\documentclass[conference]{IEEEtran}
\usepackage{graphicx} % Required for inserting images
\usepackage{cite}
\usepackage{caption}
\usepackage{amsmath}
\usepackage{subcaption}
\usepackage{xcolor}
\usepackage{algorithm}
\usepackage{algorithmic}

\IEEEoverridecommandlockouts
\begin{document}

\title{The Radon Transform, True Time Delay Beamforming, and Ultra-Wideband Antenna Arrays (Invited Paper)\\

}

\author{
\IEEEauthorblockN{Tyler Ikehara\IEEEauthorrefmark{1}, Ibrahim Pehlivan\IEEEauthorrefmark{2}, Danijela Cabric\IEEEauthorrefmark{2}, Thomas L. Marzetta\IEEEauthorrefmark{1}}
\IEEEauthorblockA{\IEEEauthorrefmark{1}A. James Clark School of Engineering, University of Maryland, College Park, MD, USA\\
Email: \{tikehara, marzetta\}@umd.edu\\
\textit{(formerly with NYU Wireless, New York University, New York, NY, USA)}}
\IEEEauthorblockA{\IEEEauthorrefmark{2}Samueli School of Engineering, University of California Los Angeles, Los Angeles, CA, USA\\
Email: ipehlivan@ucla.edu, danijela@ee.ucla.edu}
}

\date{Feburary 2025}

%\bstctlcite{IEEEexample:BSTcontrol}
\maketitle

\begin{abstract}
     The FR3 band has emerged as the major focus of 6G wireless research. FR3 cellular operation presents the challenge of extreme bandwidth combined with physically large antenna arrays. In this regime, conventional phase-shift beamforming entails a loss of coherence (beam-squint), and has to be replaced by true time delay beamforming (TTD). It happens that TTD is mathematically equivalent to taking the Radon transform of the space/time measurements. We exploit fifty years of research in the application of the Radon transform to computer tomography and to seismic exploration to elucidate the workings of TTD. We use the Radon transform combined with semblance detection and Radon slowness filtering to remove far-field signals from the measured space/time signals from a linear array, leaving only near-field signals. In turn we partition the array into sub-arrays. For each sub-array we estimate, via the semblance Radon transform,  the angles-of-arrival of the near-field signals. We then use triangulation to estimate the coordinates of the near-field sources. Finally we integrate the original space/time data along hyperbolic trajectories to extract the individual near-field signal envelopes.
\end{abstract}

\section{Introduction}
The FR3 band is of great interest as it combines the propagation advantages of lower frequency bands while having the increase in capacity that comes with higher frequency ranges\cite{5gamericas3gpptrends,miao2025surveynewmidbandfr36g,midband,baduge2025frequencyrange3isac}. This band occupies 7-24 GHz. For the first time in cellular wireless, researchers are faced with the prospect of ultra-wideband operation, with the bandwidth comparable to the carrier frequency. Conventional beamforming methods such as phase-shift beamforming aim to estimate the envelopes far-field signals which manifest themselves as plane waves, characterized by an angle of arrival. However, when dealing with ultra-wideband systems and physically large arrays, the typically used phase-shift beamforming has beam squint issues and true time delay beamforming (TTD) must be used instead. Additionally, near-field signals become more useful than far-field signals as they can be characterized both by an angle of arrival and range, allowing for a camera lens like focusing effect.

The Radon transform is the natural tool for studying TTD beamforming since the two operations are mathematically equivalent. The Radon transform is central to computer tomography \cite{Shepplogan}. Additionally, the transform has been used extensively for seismic exploration and for image processing \cite{marzettaradon,radon1,ramm2020radon,d20163d}. 

\section{Motivations}

\subsection{True Time Delay Beamforming}

Angle-of-arrival beamforming is designed to extract the envelope of a plane wave (far-field) space/time signal, which in one dimension takes the form $f(t,x) = a(t - t_0 - s_x x)$, where $s_x \in [-1/c,1/c]$ (slowness) has the units of seconds per meter, and is equal to the sine-angle divided by c. Wideband TTD adds time-shifted signals as follows: 

\begin{align}
    \hat{a}(t) = \sum_n f(t + \tau + p_x x_n, x_n) . \label{ttd}
\end{align}
When $\tau=t_0$ and $p_x = s_x$, the waveforms add coherently to yield the envelope, $a(t)$. Viewed in the frequency domain, 
\begin{align}
    \hat{A}(\omega) &= \sum_n F(\omega,x_n) e^{- i \omega \tau} e^{- i \omega p_x x_n}\\
                        &= \sum_n F(\omega,x_n) e^{- i \omega \tau} e^{- i \omega_c p_x x_n} e^{- i (\omega - \omega_c) p_x x_n}
\end{align}
where $\omega_c$ is the carrier frequency,  $|x_n| \leq\frac{L}{2}$, and $|\omega - \omega_c| \leq \pi B$. Note that $L$ and $B$ are the length of the array (m) and two-sided bandwidth (Hz) respectively. Conventional phase-shift beamforming differs from TTD in that it eliminates the term $e^{- i (\omega - \omega_c) p_x x_n}$. When the product of the bandwidth and the array-length becomes too big, there is a significant phase error in the beamforming manifested by a loss of coherence (beam-squint). Specifically, if $|(\omega - \omega_c) p_x x_n| > \frac{\pi}{4}$, or $\left (\frac{B}{f_c}\right ) \left (\frac{L}{\lambda_c}\right ) > \frac{1}{2}$, TTD must be used. Essentially, TTD is a commonly used technique when dealing with physically large arrays and wideband systems \cite{ttd1,ttd2,ttd3,ttd4}.

\begin{figure}[!t]
\centering
\captionsetup{justification=centering}
\centerline{\includegraphics[width = 0.45\textwidth]{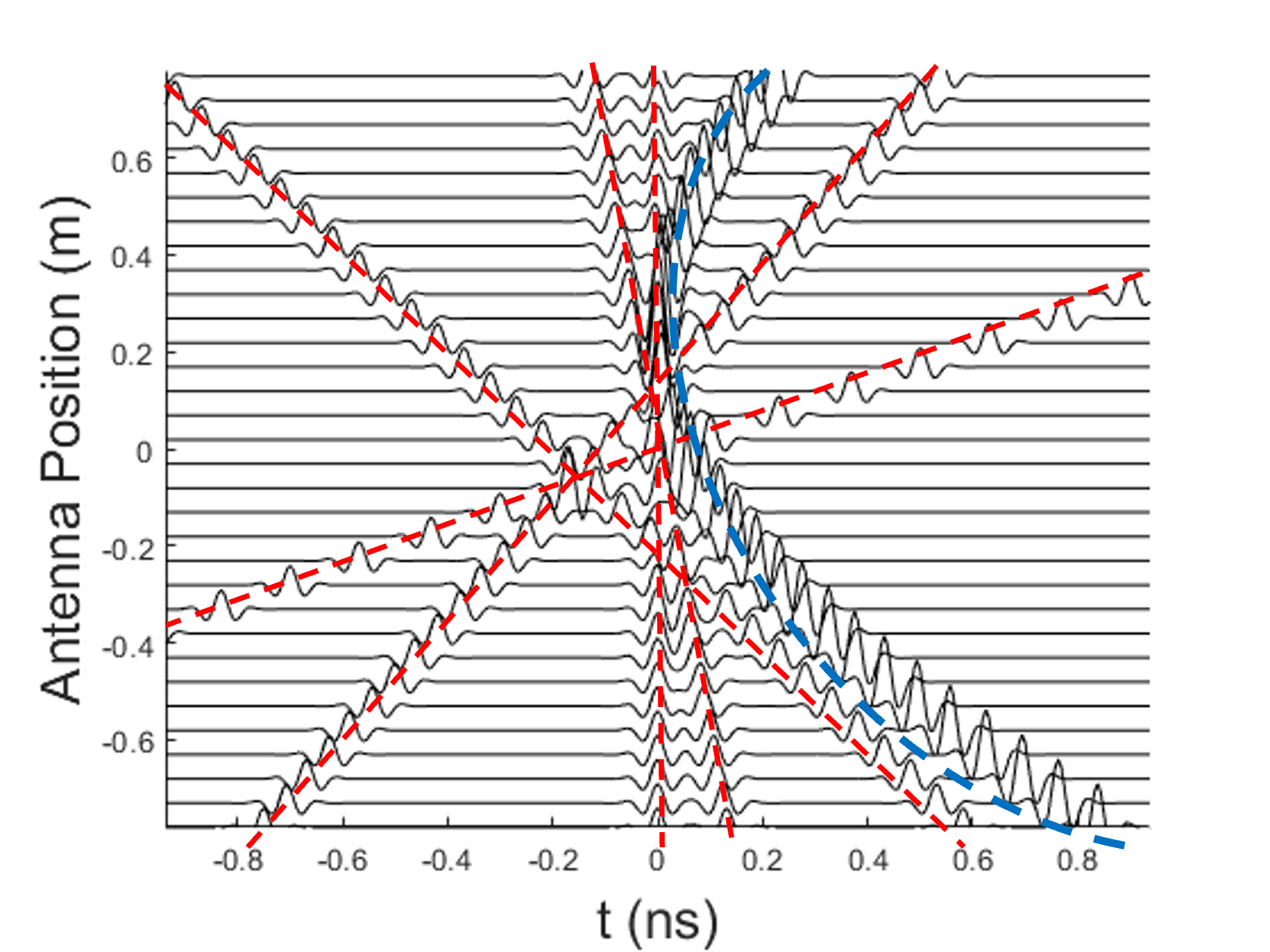}}
\caption{\raggedright A large linear array measures wideband signals from five far-field sources (straight-line space/time trajectories) and one near-field source (hyperbolic trajectory). The signals have a carrier frequency at 16 GHz with a single side bandwidth of 8 GHz and a baseband Gaussian envelope. Each trace represents the received data at each antenna in the array.}
\label{rx}
\end{figure}
\subsection{Separation of Wideband Near-field and Far-field Signals} \label{struwb}
In this research, we aim to separate the near-field signals from the far-field signals. Locally at the receive array, a far-field signal can be characterized by a plane wave, $a(t-t_0 -s_x x)$, from a source that is beyond the Fraunhofer distance from the receiver, $\frac{2L^2}{\lambda}$ \cite{10220205}. Note that $a(t)$ represents the envelope of the transmitted signals, and $s_x$ is the plane wave slowness. This means that we can only characterize the users based on their angles of arrival (AoA) as their range is ambiguous. However, with near-field signals characterized by spherical waves, $\frac{a(t - t_0 - \frac{R}{c})}{R}$, we gain the ability to localize based on AoA and also range. This focusing-like behavior is what makes it of great interest to be able to separate the near-field users from the far-field users. Additionally, this means that we can pinpoint the exact coordinate location of the near-field source.

As an example, suppose we have a FR3 wideband antenna array that is serving both near and far-field users all at unknown locations. The linear array received data is shown in Fig. \ref{rx} where each trace represents the data that the antenna in $x$ position measures. Qualitatively, we can see that the far-field arrivals look like lines and the near-field arrivals look like hyperbolas in the space/time domain. With this in mind, we will present a Radon transform based method that utilizes this behavior to separate near-field and far-field users.

\section{Proposed Method}
\begin{figure}[!t]
\centering
\captionsetup{justification=centering}
\centerline{\includegraphics[width = 0.5\textwidth]{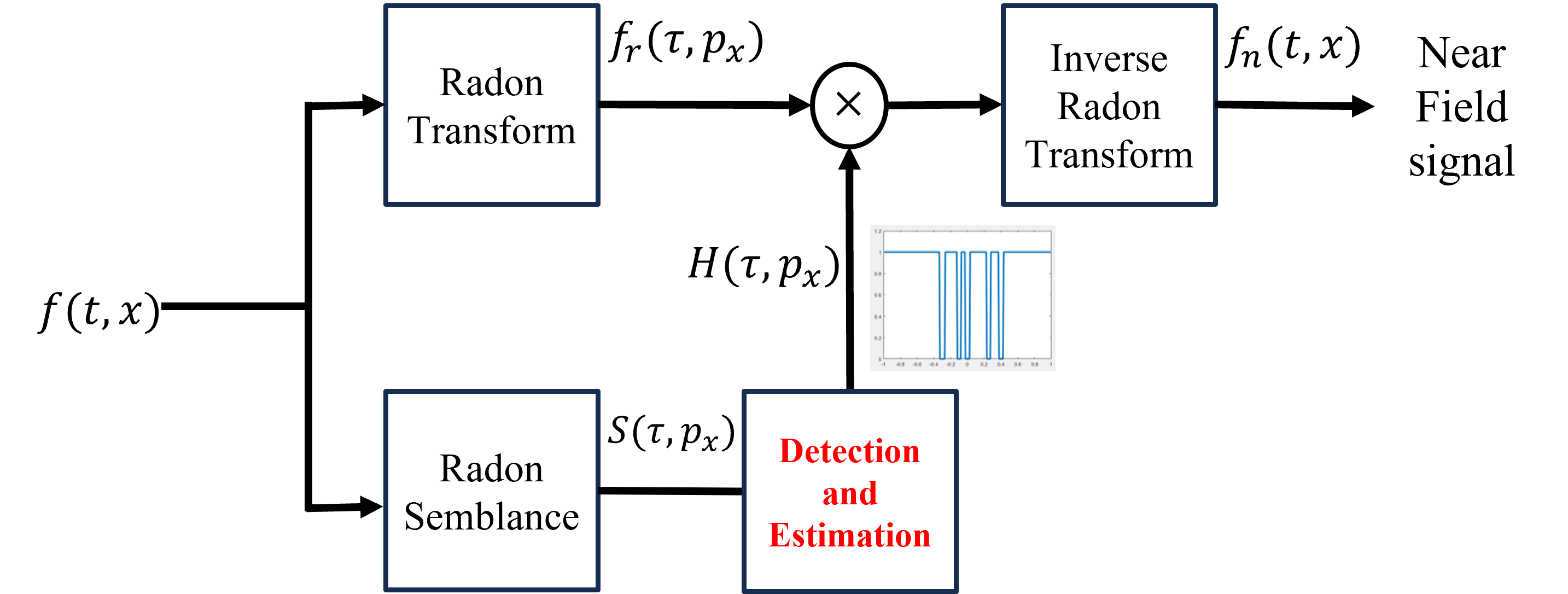}}
\caption{\raggedright Block diagram describing the proposed method of using the Radon transform to separate near-field users from far-field users. The Radon transform of the space time data is filtered against a slowness filter determined by the Radon semblance. The inverse of the product of the slowness filtered Radon transform results in the space/time data with only near-field sources.}
\label{filtering}
\end{figure}

Our Radon transform based approach to separating near-field signals from far-field signals can be seen as a block diagram in Fig. \ref{filtering}. We first take the Radon transform of the data, use the semblance to detect the slowness bands with far-field arrivals, and then use slowness filtering to eliminate the far-field users. Lastly, we take the inverse Radon transform of the filtered data to get the space/time data of only the near-field users.
\subsection{The Radon Transform}
The Radon transform is defined as:
\begin{equation}
    f_r(\tau,p_x) = \iint dxdt f(t,x)\delta(t-\tau-p_xx),
\end{equation}
where $f(t,x)$ represents the space/time data, $\tau$ is the time delay, and $p_x$ is slowness. If we integrate first over time we see that
\begin{equation}
   f_r(\tau,p_x) = \int^{L/2}_{-L/2} dx f(\tau+p_xx,x),
\end{equation}
where $L$ is the length of the linear array \cite{ttdradon}.
 Notice that if we were to discretize the Radon transform, it would be mathematically equivalent to TTD beamforming as in (\ref{ttd}). The main distinction between the two techniques is that TTD shifts the space/time data through a time delay and then combines the signals, whereas the Radon transform instead combines the signals by integrating over a multiplicity of lines in space/time parameterized by $t = \tau+p_x x$. 

 As previously mentioned in Section~\ref{struwb}, a far-field signal manifests itself in a linear array as a line in the space/time domain. If we select the $(\tau,p_x)$ pair to parameterize our integrating line such that it matches the plane wave, the signal will add coherently such as in TTD beamforming to create a peak in the $(\tau,p_x)$ domain. Of course, actual array measurements are discrete in both space and time, so we must use some sort of interpolation when taking the Radon transform.  

 Similar to the forward Radon transform, the inverse Radon transform is taken by parameterizing lines in the $(\tau,p_x)$ domain and integrating along them.
\begin{align}
    f(t,x)&=\iint  d\tau\, dp_x ~f_r(\tau,p_x)\,\delta(t-\tau-p_x x)*h_{sl}(t) \notag \\
      &=\iint d\tau\, dp_x~f_r(\tau,p_x)\,h_{sl}(t-\tau-p_x x), \\
      H_{sl}(\omega)
&= \frac{1}{2\pi}(-i\omega)\, i\, \mathrm{sgn}(\omega)
\;\;\Longleftrightarrow\;\;
h_{sl}(t)
= -\frac{1}{2\pi^{2} t^{2}}.
\end{align}
However, notice that we need to convolve with a linear filter, $h_{sl}(t)$, the Shepp-Logan filter, in order to take the Inverse Radon transform. This is due to the fact that the Radon transform is not an orthogonal transform and we need to do some sort of back-filtering in order to take the inverse \cite{radonbook}.

% \begin{equation}
%     \begin{bmatrix}
%         \boldsymbol v_t(t)\\
%         \boldsymbol v_r(t)
%     \end{bmatrix}
%     = \begin{bmatrix}
%         \boldsymbol{z_t}(t) &\boldsymbol{z_{tr}}(t) \\
%         \boldsymbol{z_{rt}}(t)& \boldsymbol{z_r}(t)
%     \end{bmatrix}
%     * \begin{bmatrix}
%         \boldsymbol{i_t}(t)\\
%         \boldsymbol{i_r}(t)
%     \end{bmatrix},
% \end{equation}

\subsection{Radon Transform of Near-field Sources}
Since the near-field sources manifest themselves as hyperbolas in space/time, it is less clear what the behavior of its Radon transform would be. Once again, we model the near-field source as a spherical waveform, $f(t,x) = \frac{a(t-t_0-\frac{R}{c})}{R}$ where $R = \sqrt{(x-x_0)^2 + (z-z_0)^2}$ and $(x_0,z_0)$ is the location of the source. Note that the Radon Transform array is assumed to be located at $\left (x\in\left [-\frac{L}{2},\frac{L}{2}\right], \,z=0 \right )$  The resulting radon transform is as follows:
\begin{align}
     f_r(\tau,p_x)  
    &= \int_{-L/2}^{L/2} dx \frac{a(\tau+p_xx-t_0-\frac{R}{c})}{R}. 
\intertext{By taking a Fourier transform we can then use the method of stationary phase to solve for the values of $(\tau,p_x)$ that correspond with the peaks of the Radon transform}
    f_r(\tau,p_x)&= \int \frac{d\omega}{2\pi}A(\omega)e^{i\omega(\tau-t_0)}\int_{-L/2}^{L/2} dx \frac{e^{i\omega(p_xx-\frac{R}{c})}}{R}\\ 
    \longrightarrow \frac{z_0^2}{c^2} &=(\tau-t_0)^2 + 2x_0(\tau-t_0)p_x + p_x^2. 
    \label{ellipse}
\end{align}
This shows that peaks of the near field arrivals are represented by ellipses in the Radon transform $\tau,~p_x$ domain. 

\begin{figure}[!t]
\centering
\captionsetup{justification=centering}
\centerline{\includegraphics[width = 0.45\textwidth]{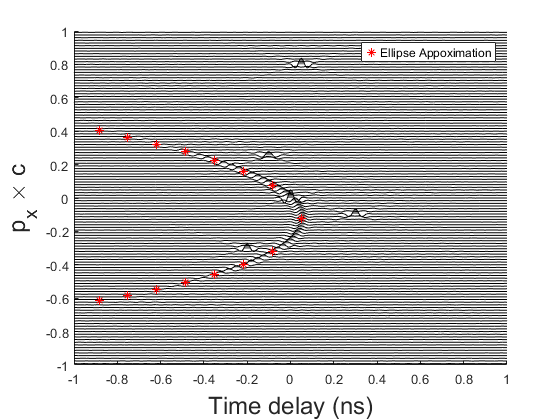}}
\caption{\raggedright Radon transform of the linear array space/time data transmitted from five far-field sources (single peaks) and one near-field source (elliptical peaks). The overlaid ellipse approximation defined by (\ref{ellipse}) matches the peaks of the near-field source.}
\label{rad}
\end{figure}

Revisiting the example from Section \ref{struwb}, the resulting Radon transform of the received data can be seen in Fig. \ref{rad}. As expected, there are five individual peaks at specific time-delay, slowness pairs that represent the far-field arrivals. Additionally, there is an elliptical curve of peaks that represent the near-field arrival. Lastly, the ellipse approximation given by (\ref{ellipse}) is shown to align with the Radon transform of the near-field signal.

\subsection{Near Field Detection: Radon Transform Semblance}
We have shown that in the Radon transform domain, far-field sources are represented by single peaks, and near-field sources are represented by ellipses. The distinction between the behavior of the users allows us to use the Radon transform to detect and classify them. However, there are potential issues in the detection if the users have Radon transforms with peaks at similar $\tau$ and $p_x$ values but are vastly different in magnitude. A well known solution is to take what is known as the semblance of the data instead \cite{marzettasemb,douzesemb}. Semblance is the ratio of coherent energy to non-coherent energy, and is normalized to have a value between zero and one.
\begin{align}
    s_w(\tau,p_x) = \frac{\left [\frac{1}{M}\sum^Mf(\tau+p_xx,x) \right]^2*w(\tau)}{\frac{1}{M}\sum^M f^2(\tau + p_xx,x)*w(\tau)},
\end{align}
where $\omega(\tau)$ is any type of window function (e.g. raised cosine, rectangular, Blackman-Harris).

The comparison of semblance with a threshold is equivalent to performing a constant false-alarm rate (CFAR) test. Semblance is equal to one if and only if the measurement comprises a pure plane wave alone. Deterministically, semblance can detect low-amplitude signals in the presence of high-amplitude signals. With this in mind, we can take the semblance of the received data and use it to detect the arrivals of plane waves and their characteristics. We are specifically interested in their slowness values as we can use them to eliminate the arrivals through filtering.
\subsection{Slowness Filtering}

The idea of slowness filtering is completely analogous to implementing a frequency bandstop filter. However, instead of eliminating specific bands of frequencies in frequency domain, we eliminate specific bands of slowness in the Radon transform domain. Specifically, frequency bandstop filtering is equivalent to multiplication of the temporal Fourier transform of the signal by a zero/one frequency response, whereas slowness bandstop filtering is equivalent to multiplication of the Radon transform of the signal by a zero/one slowness response.

To detect a plane wave arrival, we sum the semblance over time and normalize the sum such that the result is between 0 and 1. 
\begin{align}
    S(p_x) = \frac{\sum^{\tau}s_w(\tau,p_x)}{\max\left (\sum^{\tau}s_w(\tau,p_x)\right )}.
\end{align}
If $S(p_x)$ is greater than a set threshold $\epsilon$, then those values of slowness will be filtered out through a slowness bandstop filter:
\begin{align}
    f_{r_n} (\tau,p_x) &= f_r(\tau,p_x) \cdot H(\tau,p_x),\\
    H(\tau,p_x) &= 
    \begin{cases}
        0, \ \forall\, \tau &\text{if}\ S(p_x) \geq \epsilon \\
        1,  \ \forall\, \tau &\text{if}\ S(p_x) < \epsilon
    \end{cases}
    .
\end{align}

\section{Results}
\begin{figure}[!t]
\centering
\captionsetup{justification=centering}
\centerline{\includegraphics[width = 0.5\textwidth]{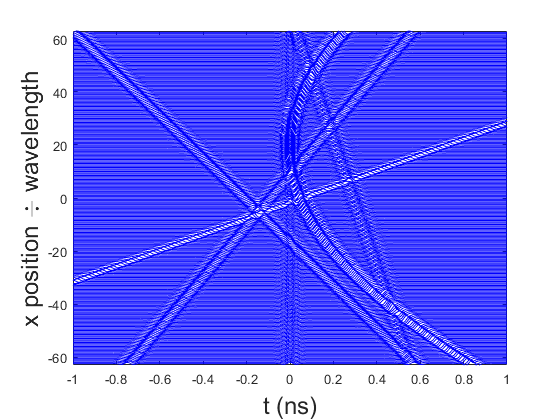}}
\caption{\raggedright A 251 element linear array receives data from five far-field sources and one near-field source. The transmitted data are wideband Gaussian pulses modulated to have a carrier frequency of 16 GHz and a single side bandwidth of 8 GHz.}
\label{rxexample}
\end{figure}

\begin{figure}[!t]
\centering
\captionsetup{justification=centering}
\centerline{\includegraphics[width = 0.5\textwidth]{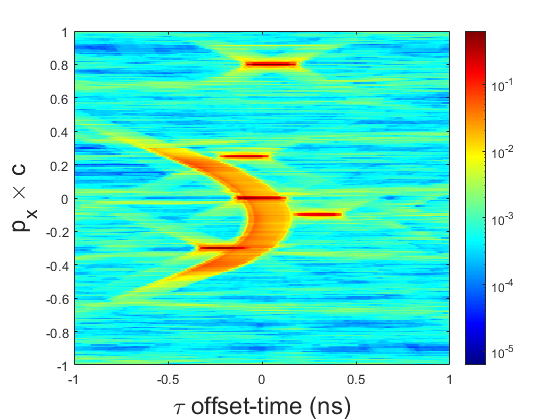}}
\caption{\raggedright Semblance of the received data using a rectangular time window with length equal to the arrival pulse duration. The red peaks correspond to a signal detection, and those closest to being equal to 1 correspond to a plane wave detection.}
\label{semb}
\end{figure}

In this section, we present the numerical results for automatic near-field user separation. For this example, we used a linear array with 251 antennas all spaced by half-wavelengths at $f_c = 24$ GHz. The transmitted data are modulated Gaussian pulses with a center frequency of $f_0 = 16$ GHz and a single side bandwidth of $B = 8$ GHz. Again, there are five far-field sources that we wish to eliminate and one near-field source that we want to keep. The received data from the linear array can be seen in Fig. \ref{rxexample} and the semblance of the data can be seen in Fig. \ref{semb}.

\begin{figure}[!t]
\centering
\captionsetup{justification=centering}
\centerline{\includegraphics[width = 0.5\textwidth]{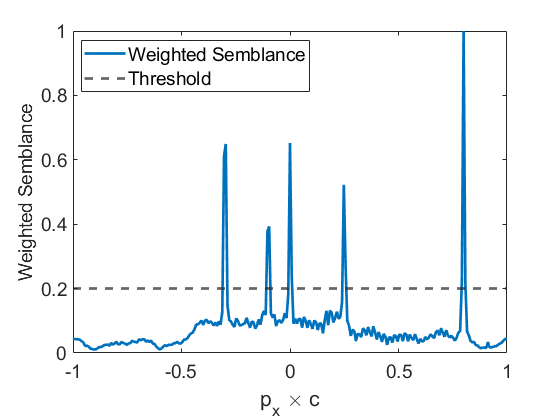}}
\caption{\raggedright Plot of weighted semblance, $S(p_x)$, as described in (\ref{sembweight}) with an arbitrary threshold $\epsilon = 0.2$. If $S(p_x)\geq\epsilon$, the slowness filter will be $H(p_x) = 0$, and all other values of $H(p_x) = 1$.}
\label{sembweight}
\end{figure}

\begin{figure}[!t]
\centering
\captionsetup{justification=centering}
\centerline{\includegraphics[width = 0.5\textwidth]{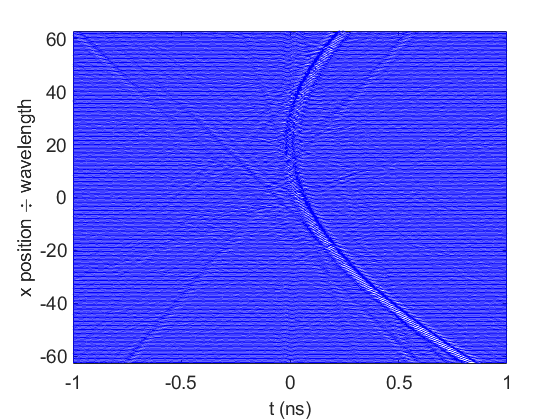}}
\caption{\raggedright Inverse Radon transform of the slowness filtered data. The near-field signal is preserved while the five far-field signals have mostly been eliminated.}
\label{invrad}
\end{figure}

 The plot of $S(p_x)$ in (\ref{sembweight}) can be seen in Fig. \ref{sembweight}. We can see that there are five spikes in the weighted semblance that correspond to the five far-field arrivals. Using this information, we then construct a slowness bandstop filter that is equal to 0 at the slowness values of the five spikes and 1 otherwise. After slowness filtering the Radon transform, the resulting inverse Radon transform can be seen in Fig. \ref{invrad}. 

It can be seen that the near-field arrival is now isolated compared to the original signal shown in Fig. \ref{rxexample}. There are still remnants of the far-field arrivals that can be seen, but they are less pronounced than before. This is likely due to the fact that our slowness filter is discontinuous and has sharp cutoffs which could lead to sidelobe interferences in space/time after taking the inverse Radon transform. We plan to further investigate these behaviors and improve our automatic detection and filtering process.


\section{Sub-Array Based Near Field Localization}
\begin{figure}[!t]
\centering
\captionsetup{justification=centering}
\centerline{\includegraphics[width = 0.5\textwidth]{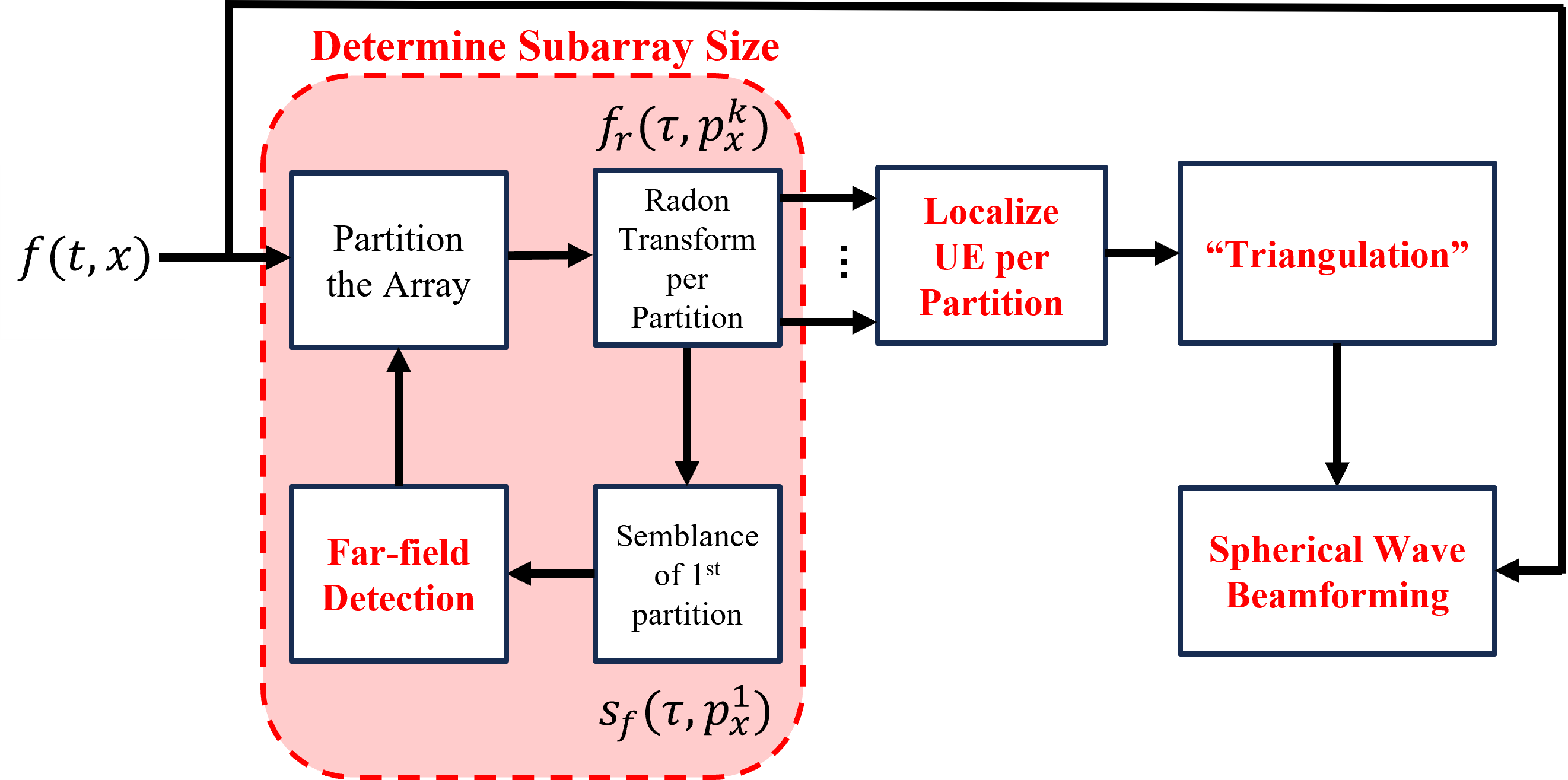}}
\caption{\raggedright Block diagram describing the sub-array based near-field localization process. The linear array is partitioned into sub-arrays with decreasing until the Semblance of the partitioned array shows a far-field detection event. A localization estimate is taken from each partition, and then we "triangulate" the UE's location using each of these estimates. Finally, spherical wave beamforming can be implemented using this fused estimate.}
\label{sub}
\end{figure}

After separating the near-field user, we can localize the user using semblance and sub-array-based triangulation. A block diagram describing this process can be seen in Fig. \ref{sub}. The near-field region boundary of an array increases with the square of the array aperture \cite{10220205}, and a user who falls within the near-field region of the entire array can be in the far-field region of a sub-array. Therefore, by partitioning the array into sub-arrays until the user falls into the far-field region of every sub-array, the user's angle to each sub-array can be independently estimated, and the user's location can be obtained by triangulation\cite{subarray_rainbow,asilomar_NF}. We can then add the signals along the hyperbolic path characterized by the user's location and see if they coherently add. 

However, this method requires prior knowledge of the user's potential location region to determine the sub-array size, limiting its practical use. Instead, we can use semblance to determine the sub-array size iteratively. More specifically, we can compute the Radon transform and semblance from a sub-array and reduce the sub-array size until the semblance indicates the user is in the far-field region. 

It is currently ongoing research for us to implement this sub-array based localization and validate it through numerical results. Additionally, a potential follow-up application would be to only use a certain amount of the sub-arrays for the localization to improve the efficiency of our system, similar to the work presented in \cite{ris}.

\section{Further Research Steps}
Going forward, there are many potential next steps for our research. Again, one of our immediate goals is to refine our automatic detection and slowness filtering techniques to eliminate the effects from the far-field sources. Additionally, in our analysis and numerics we have considered the separation of a single near-field signal from far-field signals. We wish to study if we can separate a multiplicity of near-field signals from each other or inversely, if we can eliminate the near-field arrivals to isolate the far-field users

As previously mentioned, the Radon transform is not an orthogonal transform. It is likely that implementing an orthogonalized version of the Radon transform would help to improve the results of the inverse transform routine. Also, since the near-field arrivals are hyperbolic in nature, it would be useful to be able to take the Radon transform along hyperbolas instead of lines. This would require a third parameter to characterize the hyperbolas that we will sum over, but the resulting transform would show the near-field sources more clearly.

Lastly, the numerical example that we presented was for a linear array and used Gaussian pulses as the transmitted signal. We plan on investigating the effects of instead using a longer time duration signal such as an OFDM sequence and on extending this work to planar arrays.

\section{Conclusion} 
The Radon transform is a well studied technique that has plenty of precedent that we can take inspiration from. Again, it is mathematically identical to TTD beamforming, but its alternative formulation allows us to gain new insights. In this paper, we highlight one potential use of the Radon transform in order to implement a ultra-wideband antenna array system that serves the entire FR3 range. We also showed that it is possible to separate near-field arrivals from far-field arrivals using the Radon transform semblance and slowness filtering. Lastly, we then localized the near-field user through a sub-array based approach. 

On a more philosophical note, when dealing with narrowband communication systems, we like to transform the space/time data to the frequency domain and do our analysis in it. However, when dealing with wideband systems, the behaviors of the received waveforms can become obscured upon taking the Fourier transform. Of course, this does not mean that we should neglect using FFTs when possible to speed up computational times. Nevertheless, it may be more useful to stay in the space/time domain and avoid the frequency domain, which is exactly what the Radon transform does. 
 
% \section{Numerical Radon transform Methods}
% \subsection{Wavelet Selection}

% \begin{align}
%     r(t) = \frac{1}{\sigma \sqrt{2\pi}}e^{-\frac{t^2}{2\sigma^2}}
% \end{align}

% \begin{align}
%     R(\omega) = e^{-\frac{\omega^2\sigma^2}{2}}
% \end{align}

% \begin{align}
%     &\pi B^2 A = M \\
%     &A  = \int d\omega |R(\omega)^2| = \frac{\sqrt{\pi}}{\sigma}\\
%     &M = \int d\omega |R(\omega)^2| \omega^2 = \frac{\sqrt{\pi}}{2\sigma^3} 
% \end{align}

% \begin{align}
%     B &= \frac{1}{\pi}\sqrt{\frac{M}{A}} \\
%     & =\frac{1}{\pi}\sqrt{\frac{\sqrt{\pi}/2\sigma^3}{\sqrt{\pi}/\sigma}} \\
%     &= \frac{1}{\pi\sigma\sqrt{2}} \\
%     \sigma &= \frac{1}{\pi B\sqrt{2}}
% \end{align}
% \subsection{Linear Interpolation Method}

\section{Acknowledgments}
This project was supported by funding from SpectrumX, NYU Wireless, the University of Maryland, and the University of California Los Angeles.

\bibliographystyle{IEEEtran}
\bibliography{refs}

\end{document}